# Charge spill-out and work function of few-layer graphene on SiC(0001)

O. Renault[(1)], A. M. Pascon[(2)], H. Rotella[(1)], K. Kaja[(1)*], C. Mathieu[(3)], J. E. Rault [(3)**]

P. Blaise[(1)], T. Poiroux[(1)], N. Barrett[(3)] and L. R. C. Fonseca[(2)]

[(1)] *CEA, LETI, MINATEC Campus, 17 rue des Martyrs, 38054 Grenoble Cedex 09, France.*

[(2)] *Center for Semiconductor Components, State University of Campinas, 13083-870 Campinas, SP, Brazil.*

[(3)]*CEA, DSM, IRAMIS, SPEC, 91191 Gif-sur-Yvette Cedex, France*

**Abstract**

We report on the charge spill-out and work function of epitaxial few-layer graphene on 6H-SiC(0001). Experiments from high-resolution, energy-filtered X-ray photoelectron emission microscopy (XPEEM) are combined with *ab initio* Density Functional Theory calculations using a relaxed interface model. Work function values obtained from theory and experiments are in qualitative agreement, reproducing the previously observed trend of increasing work function with each additional graphene plane. Electrons transfer at the SiC/graphene interface through a buffer layer causes an interface dipole moment which is at the origin of the graphene work function modulation. The total charge transfer is independent of the number of graphene layers, and is consistent with the constant binding energy of the SiC component of the C 1*s* core-level measured by XPEEM. Charge leakage into vacuum depends on the number of graphene layers explaining why the experimental, layer-dependent C 1*s*-graphene core-level binding energy shift does not rigidly follow that of the work function. Thus, a combination of charge transfer at the SiC/graphene interface and charge spill-out into vacuum resolves the apparent discrepancy between the experimental work function and C1*s* binding energy.

## I. INTRODUCTION

The outstanding transport properties of graphene (high carrier mobility, ballistic transport observed up to room temperature, ability to sustain large current densities, exceptional optical and mechanical characteristics) make it an attractive material for the study of two-dimensional physics, as well as for application in many devices either as passive (e.g., as electrode material) or active component (e.g., as channel material in a transistor).[1] However, to employ graphene in a device requires a suitable substrate, conserving the electrical and physical properties of free-standing graphene. If we exclude the exfoliation route for device applications, graphene can be formed from high-temperature annealing of SiC(0001) or SiC(000-1),[2-5] or directly synthesized on the surface of bulk[6] or thin[7] metallic substrates by chemical vapor deposition. Few-layer graphene (FLG) obtained on SiC(000-1) may display all the transport properties of free standing graphene,[8] however, the interface between FLG and the silicon face, SiC(0001), is more fully understood both theoretically[9] and experimentally.[10] The growth of graphene on SiC(0001) proceeds on an insulating buffer layer of a unique $(6\sqrt{3} \times 6\sqrt{3})R30°$ symmetry which decouples the subsequent graphene layers from the substrate.

For successful interface engineering and device optimization using graphene either as electrode material or active layer, control of the work function is essential as it generally governs energy level alignments through the heterostructure. This is a considerable theoretical and experimental challenge due to (i) the intrinsic low-dimensionality of FLG and the peculiar band structure of the graphene π-bands sensitive to substrate interactions and (ii) the intrinsic non-uniformity of graphene thickness in macroscopic samples.[11] The FLG work function has a layer-thickness dependency due to (i) charge transfer from electronic states at the substrate interface and (ii) charge redistribution within the FLGs by intrinsic screening. Recent work highlights how the charge transfer and charge redistribution mechanisms are sensitive to interactions of the FLG with the substrate[12,13] and between the graphene layers.[14-16]

The work function of FLG heterostructures was studied in the case of insulating,[15,16] semiconducting SiC(000-1)[11] and SiC(0001)[13,14,17,18], and various metallic substrates.[12,19,20] Ziegler *et al.* studied exfoliated FLG on SiO$_2$ and found a screening length of 4 to 5 graphene layers and work function differences of 68 meV between single layer and bilayer, and of 54 meV between bilayer and trilayer graphene.[15] Datta *et al.*[16] showed similar increase of the surface potential as a function of FLG thickness up to 5 layers on SiO$_2$ and interpreted the result in terms of intrinsic screening by the FLG of the charge transferred from a thin interfacial layer of traps at the silica surface. In the case of graphene on metals,[12] a layer dependency of the work function was also observed due to spatial variations in the charge transfer at the metal-FLG interface for domains with different in-plane orientations; moreover, contributions from metal-graphene chemical interactions to the work function are also mentioned.[12] Using X-ray Photoelectron Emission Microscopy (XPEEM), Hibino *et al.*[13] measured a 0.3 eV increase in the work function on SiC(0001) as the FLG thickness varies between 1 and 6 layers. A concomitant shift of 0.4 eV toward *lower* energies of the C 1*s* binding energy was observed, suggesting that all the electronic levels of graphene undergo a near rigid shift due to charge transfer between graphene and the SiC substrate. However, the work function and core level shifts are not perfectly anti-correlated and comparison with theoretical calculations suggests more complex electronic interactions due to chemical bonds between the buffer layer and the substrate. Thus, it is not clear whether all of the energy levels undergo a simple, band-bending like rigid shift as the number of graphene layers increases or if there are more subtle changes in the band structure.

Here, we have studied the work function of FLG epitaxially grown on 6H-SiC(0001), using high-resolution XPEEM and *ab initio* DFT calculations of a relaxed interface. Work function values obtained from theory and experiment are in qualitative agreement: the work function increase is reproduced for each additional graphene plane on the surface. Compared to free-standing graphene, the work function of the stack is modulated by the charge transferred from interface states to graphene, creating an interface dipole moment. The charge transfer is independent of the number of graphene planes but a thickness-dependent partial charge spill-out is predicted, explaining the experimental C1*s* core-level additional shift in graphene. The results provide a coherent explanation of why the layer dependent core level shift does not rigidly follow that of the work function.

## II. EXPERIMENTAL AND THEORETICAL METHODS

### A. Work function and C 1*s* core level measurements

The sample of epitaxial FLG grown on a Si-terminated 6H-SiC(0001) surface was obtained by sublimation of the SiC substrate at 1400°C for 5h under ultra-high vacuum ($10^{-6}$ Pa). This procedure resulted in micron size domains of 1, 2 and 3 LG. The vibrational fingerprint of graphene was clearly observed with Raman spectroscopy. Note that the SiC substrate was n-doped with a concentration of $10^{17}$ /cm$^3$.

Local work function measurements were performed by spectroscopic XPEEM using a *NanoESCA* instrument (Omicron NanoScience, Oxford Instruments) which has already been described elsewhere.[21] Spectroscopic XPEEM yields absolute local work function values, provided the work function of the electron analyzer is known,[13, 22] with a typical lateral resolution between 50-150 nm and an uncertainty in the measured work function of 20 meV.[23] Here, we employed soft x-ray synchrotron radiation provided by the TEMPO beamline at the SOLEIL synchrotron storage ring (Saint-Aubin, FRANCE) and 21.2 eV photons from a conventional He-discharge lamp. The combination of two excitation sources with energies well above the photoemission threshold increases the reliability of the measurement. Possible carbonaceous contamination of the graphene surface during XPEEM imaging using synchrotron radiation has been reported.[13] Therefore, the use of lower brilliance photon source, less likely to modify the surface, provides an important, independent check of the work function values.

Before XPEEM imaging, the sample was heated at 550°C in vacuum to remove adsorbates, confirmed by micro-spectroscopy of the C1*s* core level. The C 1*s* spectrum of the clean surface showed the typical graphene component at 284.4 eV and graphene-SiC interface components at higher binding energy (285.0-285.5 eV).[24] The photoemission threshold image series were recorded within two fields of views (FoVs): 34 µm (He I) and 115µm (synchrotron radiation), with a lateral resolution of 150 nm. The thickness of the FLG domains was determined by Low-Energy Electron Emission Microscopy (LEEM) and C1*s* core level XPEEM excited using synchrotron radiation (hν=400eV). The C1*s* XPEEM data were recorded with an overall energy resolution of 250 meV enabling an accurate fit of spectra from individual FLG domains using distinct core level components.

**B. Theoretical model and method**

We employed Density Functional Theory (DFT)[25] with the local density approximation (LDA) and a plane wave basis set as implemented in the ABINIT code.[26, 27] Norm-conserving pseudo-potentials were used with a plane wave cutoff of 700 eV.[28] Integration over the Brillouin zone (BZ) was performed on a 6×6×1 grid mapped according to the scheme of Monkhorst and Pack[29] carefully chosen to include the high-symmetry points characteristic of graphene. After fully relaxing a single graphene plane in vacuum (calculated lattice parameters a=b=2.458 Å, lattice angles α=β=90° and γ=60°, C–C distance = 1.426 Å), a new hexagonal eight-atom unit cell was created following Varchon *et al.*.[9] By simplifying the geometry to a $(\sqrt{3} \times \sqrt{3})R30°$ surface,[9] this interface model reduces the strain between the graphene layers and the SiC surface while maintaining the number of atoms at a practical level (53-85 atoms, depending on the number of graphene planes considered). In our interface model, the SiC slab was fully relaxed, while the graphene is stressed (tensile) in the plane to match the SiC lattice parameters. The resulting interface lattice parameters are a=b=5.487 Å, α=β=90° and γ=120°, giving 8.6% lattice mismatch along a and b directions. The c vector was chosen long enough so that the vacuum layer avoids any interaction between the system's periodic images. Moreover, the presence of the vacuum layer allowed the multilayer graphene heterostructures to relax their atomic positions along c, relieving some of the elastic energy. Convergence was achieved when the forces on the atoms were less than 0.01 eV/Å. The impact of stress on the graphene work function will be discussed below.

The resulting heterostructure containing the interface has a vacuum layer 50 Å thick to insure no sizeable cross-talk between the slab's periodic images, a 14.87 Å-thick SiC slab made of 18 silicon atoms and 18 carbon atoms, and 9 hydrogen atoms saturating the C-terminated bottom surface. Following Varchon *et al.*,[9] the SiC/graphene interface has one C atom belonging to the first graphene plane (called the buffer layer – BL) immediately on top of each Si atom at the SiC surface, except for one unpaired Si atom below the middle of the C hexagon. Other C atoms in the BL are located in intermediate positions between Si atoms below. The unpaired Si atom is important since its dangling bond plays a major role in the electronic structure of the SiC/BL system as discussed below. Mono-, bi-, tri-, and four-layer graphene (1LG, 2LG, 3LG, and 4LG) on BL/SiC were calculated (Fig. 1). For each system the thickness of the vacuum layer was kept constant at 50 Å.

The work function is the minimum energy required to extract an electron to a potential far from the surface.[30] The work function, WF, is obtained from the following expression:

$$\text{WF} = eV_{\text{vacuum}} - E_F \qquad (1)$$

where $eV_{\text{vacuum}}$ is the vacuum potential and $E_F$ is the Fermi energy. Figure 2 illustrates the process for obtaining WF theoretically, where the vacuum potential is the planar average of the total Kohn-Sham potential taken sufficiently far from the SiC/graphene slab along the direction perpendicular to the graphene surface. The SiC valence and conduction band edges (VBE and CBE) are also indicated: we obtained an indirect SiC band gap of 2.00 eV, smaller than the experimental value of 3.03 eV[31] typical of a well-known limitation of DFT/LDA.[32] Due to the presence of surface states associated with the surface Si dangling bond, the Fermi level is pinned slightly below the SiC CBE as will be detailed later (see Appendix I).

## III. RESULTS

### A. Experiment

*Graphene thickness and interface chemistry*

The FLG thickness was measured using LEEM and C 1*s* XPEEM. Figures 3(a) and 3(b) present the LEEM measurements of the thickness of individual domains. The LEEM data were obtained employing the (0, 0) specular, back-scattered electron beam. Figure 3(a) shows a typical bright-field image with a field of view of 10 μm and a lateral resolution of 30 nm. A full image series was acquired by varying E, the bias difference between the sample and electron gun, from 2 to 10 eV. Reflectivity curves of the characteristic regions are shown in Fig. 3(b). There are intensity oscillations between 1.5 and about 7.5 eV which confirm the presence of FLG domains with 1, 2 and 3 graphene layers. Following Hibino *et al.*[33], n layers of graphene give n-1 intensity minima, however, recent work by Feenstra *et al.*[34] has shown that this depends on the graphene-substrate distance. If the BL is sufficiently far from the substrate then it acts as an additional layer, thus n graphene layers will give n intensity minima in the electron reflectivity. The XPEEM data were generated pixel-by-pixel from the corresponding C1*s* image series recorded within a 17 μm field of view, in a region located in the vicinity of the one chosen for the LEEM measurements. Figure 3(c) presents a map of the intensity of the graphene component of the spectra-at-pixels, obtained after peak fitting.

The C1$s$ core level spectra of the 1, 2 and 3 LG domains marked in Fig. 3(c) are presented in Fig. 4. The energy resolution allows analysis of the chemical shifts of the individual core-level components. After Shirley background subtraction the spectra were fitted using four components related to the SiC substrate (at low binding energy), the main graphene component, and two components assigned to the buffer layer as described previously.[24] These are related to the out-of-plane C-Si covalent (S1) and in-plane sp$^2$ C-C (S2) bonding states. The main graphene component in the C1$s$ spectra shifts by 100 meV to lower binding energy with each additional layer. This will be discussed further in Sec. IV. The SiC component due to the SiC substrate has a constant binding energy of 283.7 eV. The graphene thickness is determined from the attenuation of the SiC photoemission signal by the FLG. Assuming a 0.47 nm inelastic mean-free path of C 1$s$ photoelectrons in graphite[35] with a typical 20% uncertainty, and C atom surface densities of $3.8 \times 10^{15}$ and $1.22 \times 10^{15}$ cm$^{-2}$ for graphene and SiC, respectively, we obtain FLG thicknesses of 0.51±0.05 nm, 0.69±0.07 nm, and 0.91±0.09 nm for the three domains. These values translate quite reasonably into 1-3 graphene layers, assuming a graphene interlayer spacing of 0.34 nm.

*Local work function and band shifts*

Photoemission threshold image series were recorded with synchrotron and laboratory He I radiation with FoV 115 μm and 34 μm, respectively. Prior to the fitting procedure, the images were corrected for two effects: first, the Schottky effect due to the high electrical field at the surface induced by the first extractor lens of the objective, which shifts the work function value typically by -98 meV at 12 kV extraction voltage; the second correction accounts for the non-isochromaticity of the imaging spectrometer in the dispersive direction (vertical axis on the images).[21] Figures 5(a) and 5(b) show the work function maps obtained from the photoemission threshold image series. The maps were generated by a pixel-to-pixel fit of the threshold spectra to a complementary error function.[22] This technique is much more reliable for obtaining the work function than simply extrapolating a straight line down to zero intensity in the threshold region,[13] since the theoretical shape of the onset and the energy broadening of the spectrometer are both included in the curve fitting. With this method, the uncertainty in the position of the onset obtained from the fit is ±20 meV. The histograms of the work function values extracted over the FoVs are shown in Fig. 5(c) and (d). There are clearly three distinct work function values, 4.28 ± 0.03 eV, 4.34 ± 0.03 eV, and 4.39 ± 0.03 eV, corresponding to 1, 2, and 3 LG, as measured using the C 1$s$ core level intensity. The work function therefore increases by 50-60 meV per graphene layer.

Hibino *et al.*[33] found that the work function increases by 300 meV when the FLG thickness increases from 1 to 6 LG, an average of 50 meV per layer. However, the increase was not linear, for 1-3

LG they report a 200 meV shift. Taking into account the error bars, our values are close to but slightly smaller than Hibino's. This is also the case when comparing the 1 to 2 LG work function shift with KFM results reported by Filleter et al.,[18] who found that the 2 LG increases the work function by 135 meV compared to 1 LG. For 2 LG, our value agrees particularly well with the *ab-initio* calculations of Mattausch and Pankratov.[36]

The C1*s* graphene binding energies are given in Table I. The binding energy decreases by 90 meV between 1 and 2 LG and by 80 meV between 2 and 3 LG whereas the SiC component is constant at 283.7 eV. In both cases the core level shift is significantly greater than that of the work function. Thus the C1*s* binding energy and the work function value do not undergo a rigid shift; therefore, charge transfer between the graphene and the substrate alone is not sufficient to explain the results. The theoretical calculations discussed in Sec. IV address precisely this question.

**B. THEORY**

*Graphene/SiC interface*

The interlayer distances after relaxation of the heterostructure are plotted in Fig. 6. The SiC/FLG heterostructure is also shown allowing location of the inter-planar distances. The substrate-induced corrugation of the BL is characterized by a standard deviation of 0.20 Å in the z-coordinate of the C atoms in the BL with respect to the average planar position. This gives a distribution of the distances between the BL and the SiC surface. The z-coordinate distributions for successive graphene layers become narrower. The spread in the calculated inter-planar distances is indicated by the error bars in Fig. 6. The mean BL to SiC distance is 2.27 Å with respect to the last Si layer, and 2.80 Å with respect to the last C layer, yielding an average of 2.53 Å. The SiC/BL separation of ~3.2 Å measured by Weng et al.[37] seems exceedingly large for the formation of atomic bonds and may reflect a detachment of the BL from the SiC surface upon the processing required for optical imaging of the interface. The C-Si distance for the BL C atom immediately above the Si surface atom is 1.98 Å, in excellent agreement with previous calculations.[9] The mean BL/1LG distance is 3.18 Å and the subsequent 1LG/2LG, 2LG/3LG, 3LG/4LG inter-planar distances are 3.20 Å, 3.20 Å, and 3.32 Å, respectively. These values are slightly lower than the measured graphite inter-planar distances (3.35 Å). The inter-planar distance in SiC calculated between two successive Si planes is 2.50 Å, in agreement with recent experimental results.[37]

The band structure resulting from the relaxed interface presented in the Appendix shows the characteristic Dirac cones associated with FLG graphene.

*Work function*

Table I shows the FLG work function $WF_n$ calculated using Eq. 1 for different values of n, the number of graphene planes on SiC (n=0 stands for the BL). For the bare Si-terminated SiC slab we found $WF_{SiC}$ = 3.60 eV. For SiC/BL this value is almost unchanged ($WF_0$ =3.65 eV). As the number of graphene planes increases above the BL so does the calculated work function, until 4LG, above which $WF_n$ saturates at 4.76 eV. This high WF value reflects the stress imposed on graphene in our model. Indeed, the calculated WF of bulk graphite is 4.5 eV, while for relaxed graphene it is 4.4 eV. On the other hand, the calculated WF of free-standing graphene under the 8% stress of our model is 5.1 eV. Therefore the impact of tensile stress on the WF of few-layer graphene is to increase it by ~16%. While stress tends to increase WF of several graphene planes, the deformation of the BL tends to lower its WF value: for a planar BL with only the distance to the surface Si atoms optimized, we found a larger $WF_0^{planar}$= 4.13 eV. On the other hand, we also found that for the same planar graphene model WF of the first graphene layer decreases slightly from its value obtained from the fully relaxed model, to $WF_1^{planar}$= 4.17 eV. The reason is that the first graphene layer interacts more strongly with the artificially deformed BL moving closer and raising the value of $WF_1$. For subsequent graphene layers (n=2-4), the BL is screened by the first graphene plane and does not have such a strong impact on $WF_n$.

## IV. DISCUSSION

Here we discuss the consistency of our interface model results and the experimental data on the work function and the C 1$s$ core levels. The experimental work function shifts by 50-60 meV per graphene layer, whereas the C 1$s$ binding energy in graphene shifts by almost 100 meV per layer. Figure 4 shows that the interface chemistry, represented by the S1 and S2 components of the local C 1$s$ core level spectra, does not change with the number of graphene layers on SiC. Therefore the increase in the SiC/FLG work function with the number of graphene layers should be of an electrostatic origin, related to an increase of the interface dipole. To confirm this hypothesis we have calculated the interface dipoles $D$ by integrating the net charge density $\Delta\rho$, defined as the difference between the heterostructure planar averaged (along x and y) charge density and the SiC and graphene bulk planar averaged charge densities, multiplied by the displacement vector d:

$$D = \frac{1}{A} \int_{-\infty}^{+\infty} d(z)\Delta\rho(z)dz, \qquad (2)$$

where A is the interface area and the z direction is perpendicular to the interface. To estimate the interface dipole and how it is affected by the number of graphene layers on SiC, the net charge density was obtained using the following two methods.

In method I the net charge was obtained by subtracting the charges of SiC and graphene slabs from the heterostructure charge. If the impact of the interface on the location of the atoms is short-ranged, quickly decaying into the bulk-like regions, then this procedure is able to remove the bulk charges from both sides of the interface, leaving only the net interface charge. Formally the net charge density of the interface structure, $\Delta\rho^{SiC+FLG}$, can be written as

$$\Delta\rho^{SiC+FLG} = \rho^{SiC+FLG} - \rho^{SiC} - \rho^{FLG} = (\rho^{\overline{SiC}} - \rho^{SiC}) + (\rho^{\overline{FLG}} - \rho^{FLG}), \quad (3)$$

where $\rho^{SiC+FLG}$ is the total charge density of the interface structure, $\rho^{SiC}$ and $\rho^{FLG}$ are the total charge densities of the SiC and FLG slabs, respectively, and $\rho^{\overline{SiC}}$ and $\rho^{\overline{FLG}}$ are the total charge densities of the SiC and FLG sides of the interface which include the charge exchanged across the interface. The total charge densities can be further decomposed in their positive and negative components arising from the contributions from the ions and electrons, respectively. Therefore

$$\Delta\rho^{SiC+FLG} = [(\rho_{ion}^{\overline{SiC}} - \rho_{ion}^{SiC}) + (\rho_e^{\overline{SiC}} - \rho_e^{SiC})] + [(\rho_{ion}^{\overline{FLG}} - \rho_{ion}^{FLG}) + (\rho_e^{\overline{FLG}} - \rho_e^{FLG})]. \quad (4)$$

The alignment between the charge densities of the SiC of the heterostructure and the bare SiC slab was straightforward as shown in Figure 7, since the interface has little effect on the nearby SiC layers. For this reason, the first term of the left bracket in Eq. 4, $\rho_{ion}^{\overline{SiC}} - \rho_{ion}^{SiC}$, can be neglected. In the case of the graphene slab, the alignment with the graphene in the SiC/graphene heterostructure is more difficult because the interface has a considerable impact on the graphene layer separation, moving them closer to each other than in the free-standing graphene slab. In other words, the contribution of the ionic charge density in graphene shown in Eq. 4, $\rho_{ion}^{\overline{FLG}} - \rho_{ion}^{FLG}$, is considerable and extends several angstroms away from the interface. We assumed the most external graphene layer in the heterostructure to be the reference for alignment between the charge densities of the 5-layer graphene slab (5LG) and the SiC/BL/4LG heterostructure, since the 5[th] graphene layer is the layer least affected by the interface in all our stack models containing different number of graphene planes. From that alignment we obtained the $z_0 = 0$ position of the graphene plane in 5LG closest to SiC. For the other alignments (between the (n+1)LG slabs and SiC/BL/nLG heterostructures, for n = 0, 1, 2, 3) the graphene plane closest to SiC was fixed at the value of $z_0$ determined from the 5LG. The upper panel of Fig. 7 shows the resulting mismatch between the position of the charge density peaks in the heterostructure and in the graphene slab obtained with method I. Notice that as the number of graphene layers increases the mismatch decreases for the most external layers, as expected. The lower panel in Fig. 7 shows the net charge density thus obtained. As the number of graphene layers increases the spread of the net charge density remains almost constant in SiC but broadens in graphene, with an exponential decay exp(-k/z), where k = 0.11Å. Despite the intuitive appeal of method I, the difficulty in aligning the graphene slab with the graphene part of the heterostructure raises the question wether the role played by the graphene ionic contribution masks the effect of the smaller net charge density. To verify that this limitation is not significant for our conclusions, we have employed a second approach to calculate the net charge density.

In method II we have subtracted from the density of the graphene side of the interface the densities of the appropriate number of single graphene planes. In this case, $\rho_{ion}^{FLG}$ and $\rho_e^{FLG}$ in Eq. 4 are the ionic and electronic charge densities of a free-standing single graphene plane repeated the necessary number of times depending on the number of graphene planes in the SiC/FLG structure. Although this method does not take account of the electronic charge located in the inter-planar regions of graphene in the stack, it intrinsically accounts for the inter-planar graphene separation. The total charge alignments for SiC/BL, SiC/BL/1LG and SiC/BL/4LG obtained with method II are shown in the upper panel of Fig. 7. The charge densities are better aligned, implying that in this case both $\rho_{ion}^{\overline{SiC}} - \rho_{ion}^{SiC}$ and $\rho_{ion}^{\overline{FLG}} - \rho_{ion}^{FLG}$ in Eq. 4 can be neglected. The resulting net charge density, which is one order of magnitude lower than that obtained using method I, is shown in the lower panel of Fig. 7. The results of method II are given in parentheses in Table I.

The results obtained with both methods follow the same trends. Table I shows the dipole per unit area, obtained from Eq. 2 with the lower and upper integration limits replaced by the middle of the SiC slab and the vacuum region beyond the last graphene plane, respectively. The dipole increases with the number of graphene layers for the cases SiC/BL up to SiC/BL/4LG. To simplify the following discussion, we only show results obtained with method I.

The interface dipole variation supports the hypothesis that the WF dependence on the number of graphene planes is due only to electrostatics, *i.e.* interface charge transfer to an increasing number of graphene layers. However it does not explain the simultaneous near invariance of the measured C1*s* core level in SiC and the amplitude of the C 1*s* core level shift in graphene with respect to the WF shift.

The interface dipole is the product of the charge transferred across the interface and the separation of the positive and negative charges. Table I shows that the calculated electronic charge transferred from SiC to graphene is almost constant as the number of graphene layers increases. On the other hand, the spread of the charge transferred at the graphene side increases with the number of planes as shown in the lower panel of Fig. 7. Therefore, the SiC/graphene dipole only changes with the number of graphene layers because the more graphene planes there are the larger is the separation between the charge centroids at each side of the interface, with no impact on the electronic states on the SiC side of the interface. This explains why the C1*s* level in SiC is unchanged.

However, due to the atomically-thin character of few-layer graphene, part of the transferred charge leaks into vacuum as shown in Table I. The fraction of the leaked charge originating from the transferred charge (*i.e.*, excluding the leaked charge belonging to the topmost graphene plane) was calculated by subtracting the charge of the graphene slab from that of the interface structure. The latter is obtained by integrating the net charge density starting from the average z-coordinate of the topmost graphene plane to some distance a few angstroms away in the vacuum region. The convergence of the integration was checked by changing the endpoint of the integral in vacuum. As the number of graphene layers increases, the amount of leaked charge calculated with this method decreases from 80% of the transferred charge for SiC/BL to 14% for SiC/BL/4LG. In other words, thicker graphene slabs are able to contain more negative charge, shifting the C1$s$ core level in graphene to lower binding energy, as observed experimentally.

These results should be of interest in the perspective of more complex heterostructures, for example by epitaxial growth of a semiconductor on top of FLG to open and control gaps in the graphene. The layer dependent charge spillage may therefore be as important a parameter as the work function shift, and the combination of the two may allow more complex band engineering.

**CONCLUSION**

We have used high-resolution energy-filtered XPEEM and *ab initio* DFT calculations of a relaxed interface model to determine the work function of few layer graphene on SiC(0001). Work function values obtained from theory and experiments are found in qualitative agreement with an increase of the work function with each additional graphene layer. Compared to isolated graphene the work function is modulated by the charge transfer from interface states to graphene, creating an interface dipole moment. In the calculations, the charge transferred is independent of the number of graphene planes, consistent with the constant C 1$s$-SiC core-level binding energy measured by XPEEM. A layer-dependent partial spill-out of the transferred charge is predicted theoretically and explains the experimental C 1$s$-graphene core-level binding energy variations, with a layer-dependent shift not rigidly following the work function evolution.

**Acknowledgments**

The theoretical work was financially supported by the Nanosciences Foundation of Grenoble (France) in the frame of the Chairs of Excellence awarded to L.R.C.F. A.M.P. and L.R.C.F. also thank INCT/Namitec and CNPq for financial support. The experimental part was performed on the Nanocharacterization Platform of MINATEC and supported by the GRAND project and the French National Research Agency (ANR) through the Recherche Technologique de Base (RTB) program. J.R. was funded by a CEA CFR grant and CM was funded by the CEA Nanosciences project k-PEEM.

APPENDIX I

*Band structure of the modeled SiC/FLG heterostructures*

Figure A1 shows the total density of states (DOS) projected on particular Si atoms of the SiC slab, and on C atoms of the BL. The well-defined SiC band gap in the inner part of the SiC slab (2.0 eV) taken on the Si atom #4 in the figure is slightly smaller than its calculated bulk value (2.32 eV) due to the small thickness of the SiC slab employed, which allows surface and interface states to reach the central region of the slab. The PDOS are similar for all C atoms in the buffer layer, which are represented by only one line in the graph. The PDOS are also similar for the bonded Si interface atoms #1a and #1b, both represented by only one line in the graph. The PDOS associated with the lonely Si interface atom #2 differs from the ones taken at atoms #1a and #1b only by its higher density of states near the edges of the band gap. This is expected since those edge states are associated with the dangling bond localized at the unpaired interface atom #2. The edge state near the SiC CBE is fully occupied and pins the Fermi level near the SiC CBE. Even though these interface states are quite close to the SiC band edges, conferring a metallic character to the interface as previously observed,[38] they are rather localized with very little dispersion indicating that the interface is a poorly conductive metal.

Figure A2 shows the band structure obtained from the DFT calculations for the SiC/BL system, SiC/BL/1LG and SiC/BL/2LG. The most important feature is the absence of cones at the K high symmetry point in the Brillouin zone for the BL whereas the Dirac cones are present for both the 1LG and 2LG systems. The absence of cones shows that the BL does not have the electronic structure of an isolated 1LG but behaves as a transition region between SiC and graphene.[38] Figure A2(b) also shows that the addition of graphene layers on SiC/BL does not change the characteristics of the interface state associated with the lonely Si atom. For SiC/BL/1LG the cone at K is recovered, with the Fermi level pinned 0.6 eV above the Dirac point, indicating electron transfer from SiC to graphene. For BL+2LG the two $\pi$ and $\pi^*$ bands typically associated with two graphene planes can be easily identified, showing a small band gap of 0.2 eV. The cones indicate that all graphene planes separated from SiC by the BL are structurally intact. A crucial point, however, is that the destruction of the Dirac cones in the BL does not result from its deformation, but instead it is induced by the SiC surface states. The distinction between the impact of geometrical deformation and surface states on the BL was achieved by calculating the band structure of the deformed BL in the absence of the SiC slab. The result is shown in the inset of Fig. A2(a), where the Dirac cone appears at K as in unbuckled graphene.

**Table captions**

TABLE I: Calculated and measured work functions, measured core level binding energies, calculated interface dipoles, calculated charge transferred from SiC to FLG, and calculated charge spill-out. The fourth row was calculated using methods I and II (in parenthesis) described in the text.

**Figure captions:**

FIG. 1. (Color online) Interface models of multiple layer graphene (nLG) on SiC in the presence of a buffer layer (BL): (a) SiC/BL; (b) SiC/BL/1LG; (c) SiC/BL/4LG. Yellow (big) balls: Si; gray (medium) balls: C; white (small) balls: H.

FIG. 2. (Color online) Planar averaged potential for the SiC/BL slab . $E_F$, CBE, VBE, and WF are the calculated Fermi energy, SiC conduction and valence band edges, and the work function, respectively. The buffer layer (BL) is located at the x-axis origin.

FIG. 3. (Color online) (a) Typical 10 µm bright field LEEM image; (b) electron reflectivity curves extracted from the areas marked in (a); (c) XPEEM C 1$s$ map (FoV: 17 µm, hv=400eV) of the graphene component intensity.

FIG. 4. (Color online) High resolution local C 1$s$ core-level of the FLG domains marked in Fig. 3(c). The main graphene peak is dark grey, the interface components S1, S2, are in lighter grey and the SiC component black. The vertical line indicates the unchanging position of the SiC component.

FIG. 5. (Color online) Work function maps and corresponding distribution of domains obtained from spectroscopic XPEEM images at threshold, in the case of (a) synchrotron (FoV: 115µm) and (b) He I excitation (FoV: 35µm). The respective histograms of the work function values across the whole field of view are shown in (c) and (d).

FIG. 6. (Color online) Calculated (triangles) and measured[37] (squares) average inter-planar distances between carbon planes indicated in the model below. The circle is the calculated average interplanar distance between the buffer layer and the Si plane at the interface. The calculated error bars were obtained from the farthest and closest atoms along z belonging to neighboring atomic planes. The calculated error bars are close to zero at the middle of the SiC slab. Gray: C; orange (dark gray): Si; white: H.

FIG. 7. (Color online) (a) SiC/BL; (b) SiC/BL/1LG; (c) SiC/BL/4LG. Top panel:: charge density alignment (black: SiC/graphene; blue dotted: SiC slab) using different methods (see text): red dashed: graphene slab (method I); green dot-dashed: isolated graphene planes (method II). Bottom panel: net charge densities (solid: subtraction between the SiC/graphene density and the SiC slab and graphene slab densities; red dashed (magnitude multiplied by 8): subtraction between the SiC/graphene and the SiC slab and isolated graphene planes densities).

FIG. A1. (Color online) Projected density of states (PDOS, left) taken at the atoms indicated in the interface model (right). Gray: C; orange (dark gray): Si; white: H.

FIG. A2. (Color online) Band structure for the relaxed interface system: (a) SiC/BL (inset: distorted BL without the presence of the SiC slab), (b) SiC/BL/1LG and (c) SiC/BL/2LG. The Fermi level is the horizontal dotted line at 0 eV. The band gap state (red dashed line) belongs to the unpaired (lonely) Si atom at the SiC/BL interface. The band structures in (b) and (c) show the linear and double parabolic dispersions (highlighted in the squares), characteristic of freestanding monolayer and bilayer graphene.

**Table I**

|  | SiC slab | BL | 1LG | 2LG | 3LG | 4LG |
|---|---|---|---|---|---|---|
| **$WF_n$ (eV), calculated** | 3.60 | 3.65 | 4.31 | 4.55 | 4.75 | 4.76 |
| **$WF_n$ (eV), measured** | ---- | ---- | 4.28 ± 0.03 | 4.34 ± 0.03 | 4.39± 0.03 | ---- |
| **C 1$s$ BE (eV)** | 283.7 |  | 284.57± 0.05 | 284.48± 0.05 | 284.40± 0.05 |  |
| **Interface dipole (eV)** | ---- | 0.14 (0.007) | 0.15 (0.007) | 0.19 (0.009) | 0.20 (0.010) | 0.22 (0.012) |
| **Charge transferred (e/Å$^2$)** | ---- | 0.7 | 0.59 | 0.55 | 0.54 | 0.58 |
| **Charge spill-out (e/Å$^2$)** | ---- | 0.56 | 0.43 | 0.32 | 0.20 | 0.08 |

**Figure 1**

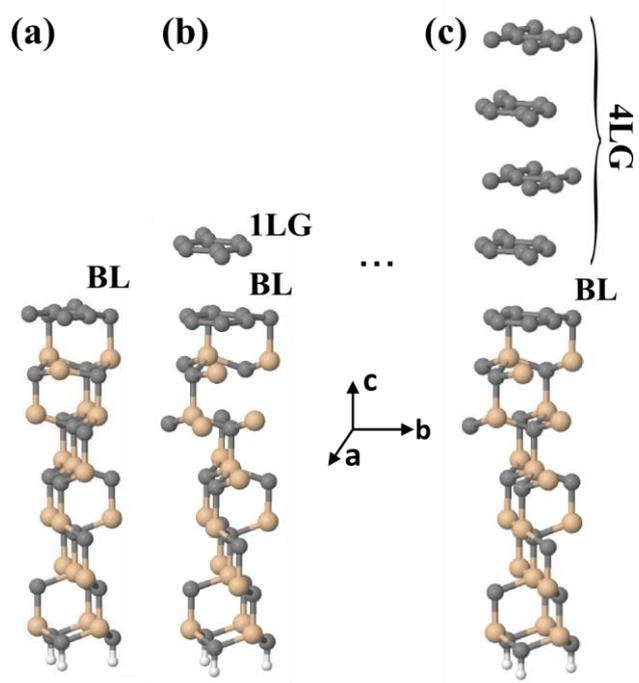

**Figure 2**

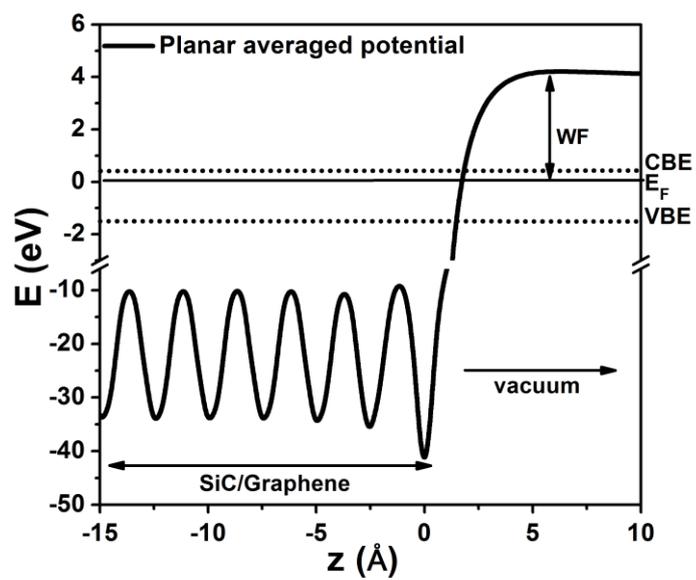

**Figure 3**

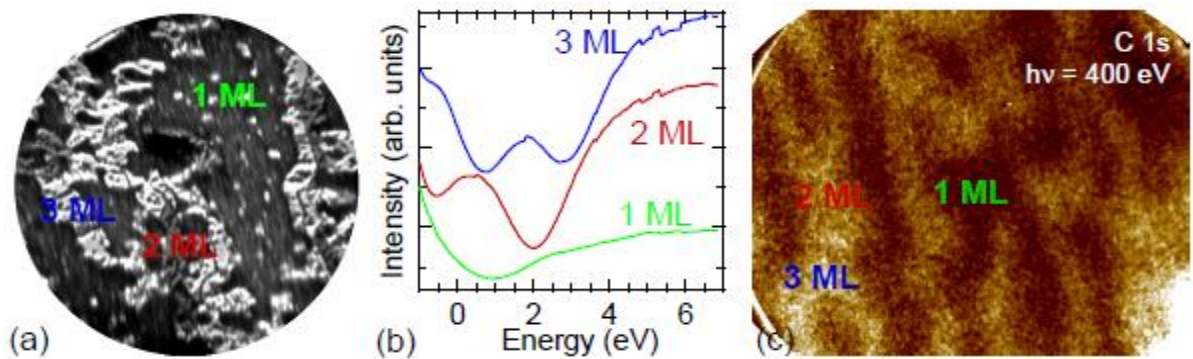

**Figure 4**

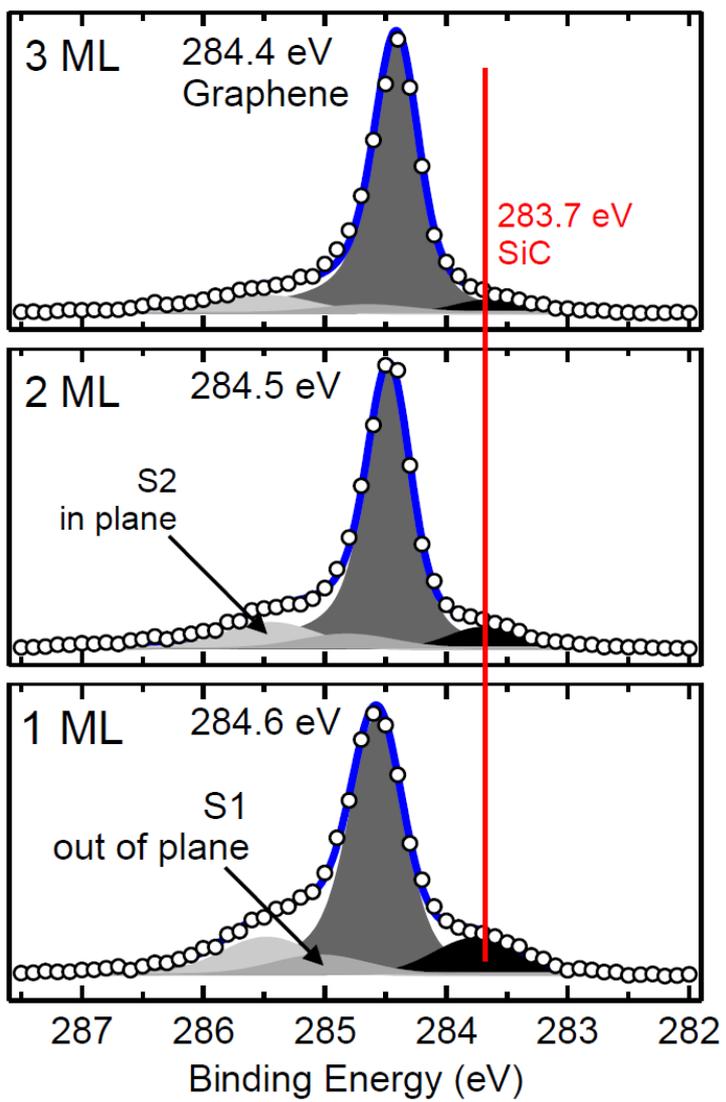

**Figure 5**

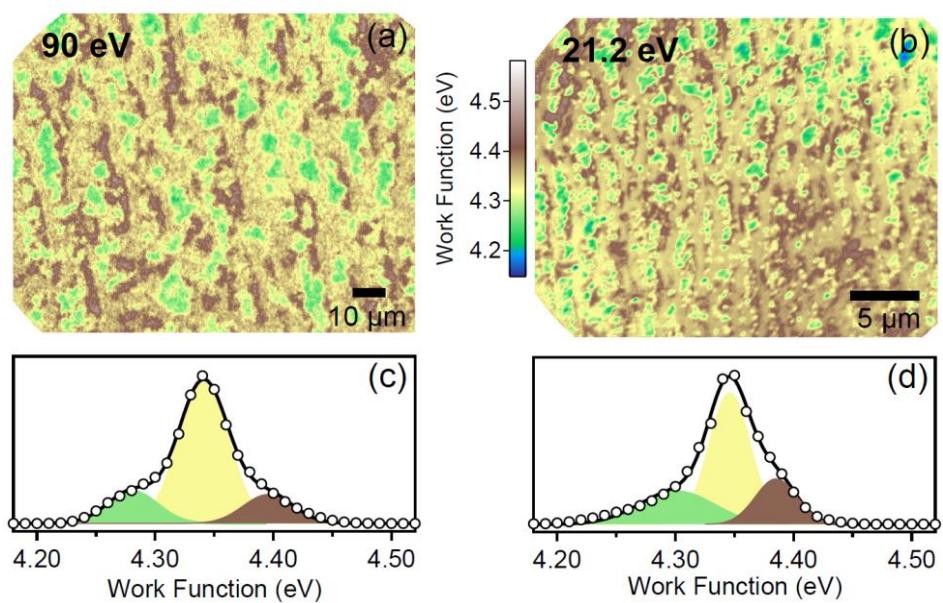

**Figure 6**

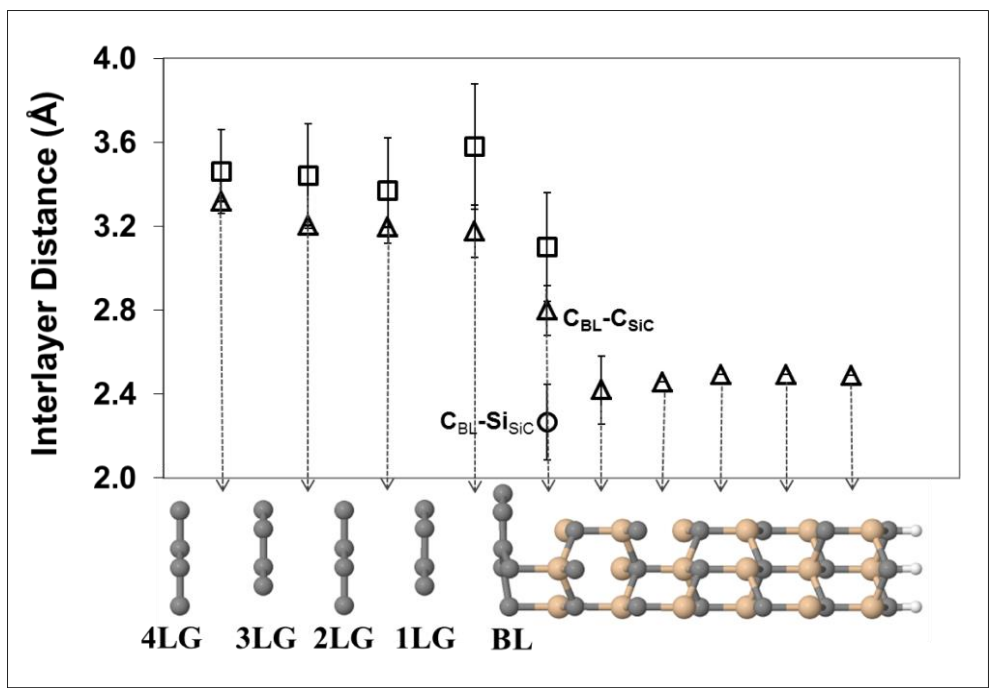

**Figure 7**

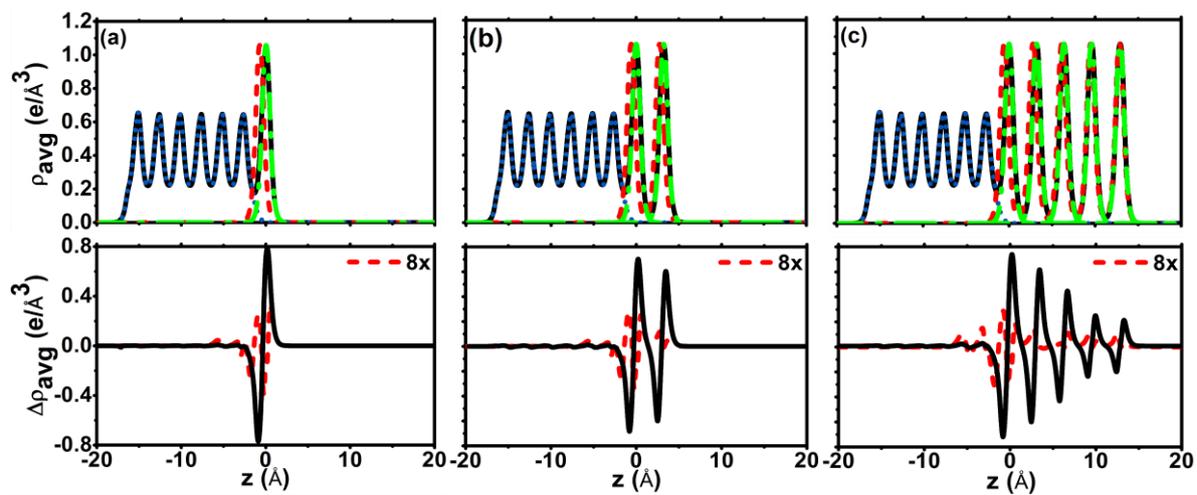

**Figure A1**

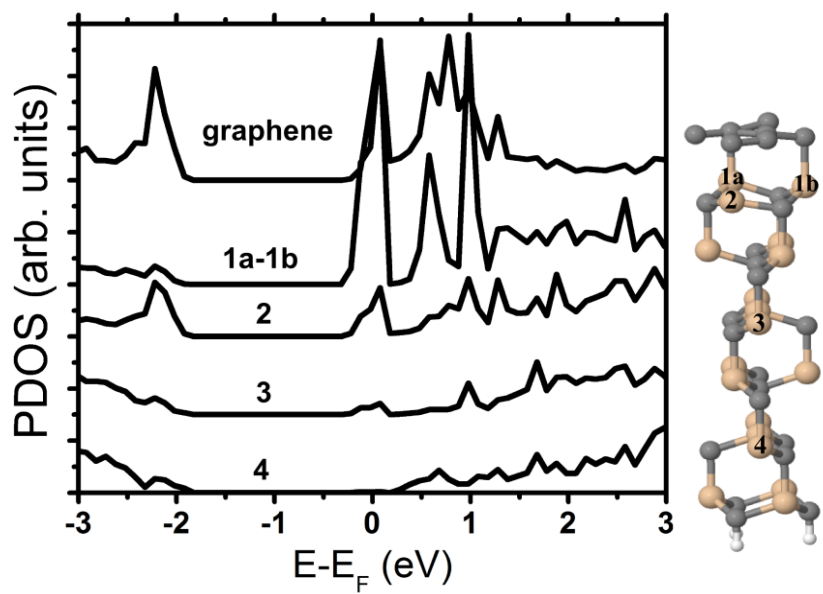

**Figure A2**

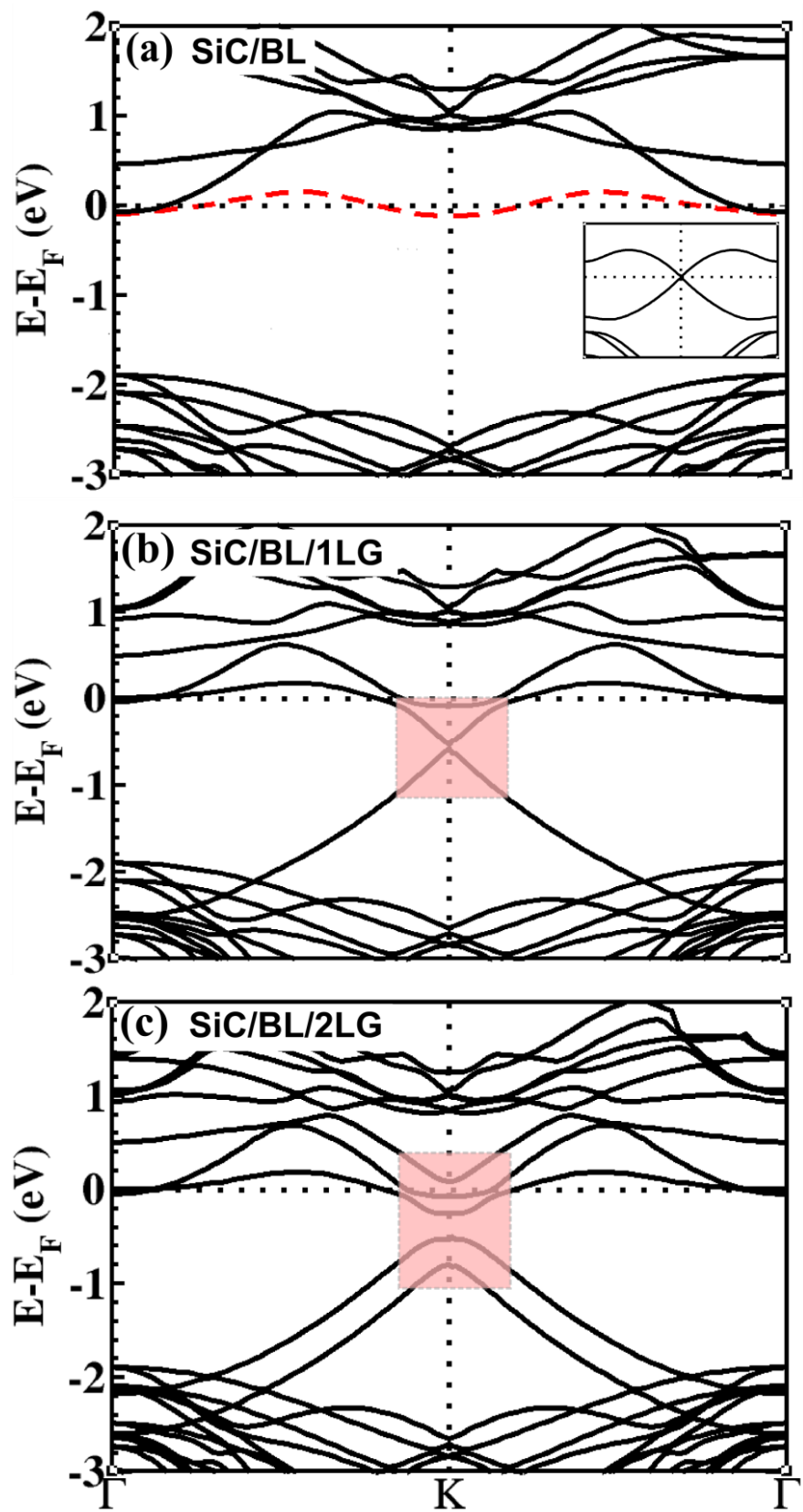